\documentclass[twocolumn]{aastex61}
\pdfoutput=1 
\usepackage{amsmath,amstext}
\usepackage[T1]{fontenc}
\usepackage{apjfonts} 
\usepackage[figure,figure*]{hypcap}
\usepackage{natbib}
\usepackage{color}

\shorttitle{Scattering Size Distribution}
\shortauthors{Lawler et al.}

\begin{document}

\title{OSSOS. VIII. The Transition Between Two Size Distribution Slopes in the Scattering Disk}

\author[0000-0001-5368-386X]{S.~M. Lawler}
\affiliation{NRC-Herzberg Astronomy and Astrophysics, National Research Council of Canada, Victoria, BC, Canada}
\author[0000-0001-7032-5255]{C. Shankman}
\affiliation{Department of Physics and Astronomy, University of Victoria, Victoria, BC, Canada}
\affiliation{NRC-Herzberg Astronomy and Astrophysics, National Research Council of Canada, Victoria, BC, Canada}
\affiliation{City of Toronto, Toronto, ON, Canada}
\author[0000-0001-7032-5255]{JJ. Kavelaars}
\affiliation{NRC-Herzberg Astronomy and Astrophysics, National Research Council of Canada, Victoria, BC, Canada}
\affiliation{Department of Physics and Astronomy, University of Victoria, Victoria, BC, Canada}
\author[0000-0003-4143-8589]{M. Alexandersen}
\affiliation{Institute of Astronomy and Astrophysics, Academia Sinica, Taipei, Taiwan}
\author[0000-0003-3257-4490]{M.~T. Bannister}
\affiliation{Astrophysics Research Centre, Queen's University Belfast, Belfast, UK}
\author[0000-0001-7244-6069]{Ying-Tung Chen}
\affiliation{Institute of Astronomy and Astrophysics, Academia Sinica, Taipei, Taiwan}
\author{B. Gladman}
\affiliation{Department of Physics and Astronomy, The University of British Columbia, Vancouver, BC, Canada}
\author{W.~C. Fraser}
\affiliation{Astrophysics Research Centre, Queen's University Belfast, Belfast, UK}
\author{S. Gwyn}
\affiliation{NRC-Herzberg Astronomy and Astrophysics, National Research Council of Canada, Victoria, BC, Canada}
\author{N. Kaib}
\affiliation{HL Dodge Department of Physics \& Astronomy, University of Oklahoma, Norman, OK, USA}
\author[0000-0003-0407-2266]{J.-M. Petit}
\affiliation{Institut UTINAM, CNRS-Universit\'e de Franche-Comt\'e, Besan\c{c}on, France}
\author[0000-0001-8736-236X]{K. Volk}
\affiliation{Lunar and Planetary Laboratory, University of Arizona, Tucson, AZ, USA}

\begin{abstract}
The scattering trans-Neptunian Objects (TNOs) can be measured to smaller sizes than any other distant small-body population.
We use the largest sample yet obtained, 68 discoveries, primarily by the Outer Solar System Origins Survey (OSSOS), to constrain the slope of its luminosity distribution, with sensitivity to much fainter absolute $H$ magnitudes than previous work. 
Using the analysis technique in \citet{Shankmanetal2016}, we confirm that a single slope for the $H$-distribution is not an accurate representation of the scattering TNOs and Centaurs, and that a break in the distribution is required, in support of previous conclusions. 
A bright-end slope of $\alpha_b=0.9$ transitioning to a faint-end slope $\alpha_f$ of 0.4-0.5 with a differential number contrast $c$ from 1 (a knee) to 10 (a divot) provides an acceptable match to our data.
We find that break magnitudes $H_b$ of 7.7 and 8.3, values both previously suggested for dynamically hot Kuiper belt populations, are equally non-rejectable for a range of $\alpha_f$ and $c$ in our statistical analysis. 
Our preferred divot $H$-distribution transitions to $\alpha_f=0.5$ with a divot of contrast $c=3$ at $H_b=8.3$, while our preferred knee $H$-distribution transitions to $\alpha_f=0.4$ at $H_b=7.7$.
The intrinsic population of scattering TNOs required to match the OSSOS detections is  $3\times10^6$ for $H_r<12$, and $9\times10^4$ for $H_r<8.66$ ($D\gtrsim100$~km), with Centaurs having an intrinsic population two orders of magnitude smaller.

\end{abstract}


\section{Introduction}

The populations of small bodies in our Solar System are incrementally grinding themselves into dust through mutual collisions.  
On short timescales, collisions are infrequent, though on occasion the aftermath can be directly observed \citep[e.g.][]{Jewittetal2010}. 
Over the age of the Solar System, collisions may be the main force that shaped the observed size distribution of all but the largest trans-Neptunian objects (TNOs) \citep{Schlichtingetal2013}, or the size distribution may be a result of formation conditions \citep{Fraseretal2014}.
As dynamical evolution is not size-dependant for these small TNOs, we don't expect the size distribution to be affected by removal of TNOs from the scattering population due to interactions with the giant planets.
The size distribution of populations that are shaped by collisions can be described by a power law of the form $\frac{dN}{dD}\propto D^{-q}$, where an idealized infinite collisional cascade will produce an exponent of $q=3.5$ \citep{Dohnanyi1969}.

In the outer Solar System, the luminosity distribution must be used as a proxy for the size distribution, because TNOs are unresolved.
Luminosity is measured as an apparent magnitude, which can be directly converted to an absolute magnitude $H$ when combined with a measured distance.
$H$ magnitude can then be directly mapped to diameter, as long as an albedo is measured (or assumed).  
A handful of small ($H\sim9-14$) TNOs and Centaurs have had their albedos measured observationally, and they range from 4-16\% \citep{Duffardetal2014}. 
The size distribution can be written in terms of absolute magnitude $H$ as $\frac{dN}{dH}\propto 10^{\alpha H}$, where the size distribution exponent $q$ is related to the $H$-magnitude exponent $\alpha$ by $q=5\alpha+1$ \citep[assuming albedo is size-independent;][]{Irwinetal1995,Petitetal2008,Fraseretal2008}.

Measuring the size distribution of a small body population tells us about their composition, collisional processes that shape them, and may also provide information on their formation.  
Collisional simulations of the asteroid belt \citep[e.g.][]{Bottkeetal2005,PanSchlichting2012} have found that the sizes of the largest asteroids are set by the initial formation sizes,
which in combination with mass depletion of the asteroid belt (caused by Jupiter's migration), sets any structure in the size distribution.
The size distribution of the asteroids can be measured to much smaller sizes (larger $H$-magnitudes) than the TNOs due to the fact that it is much closer and thus smaller objects will be above survey detection limits.  
The asteroid size distribution at smaller sizes shows intriguing structure, which collisional simulations have shown to likely be caused by a combination of formation size and the initial number density of the asteroid belt; the transition between primordial and collisionally evolved populations happens at $\sim10-100$~km \citep{Bottkeetal2005,Morbidellietal2009bornbig}.
By measuring the size distribution of the Kuiper Belt across several orders of magnitude in size, as has been done in the asteroid belt, we may gain an additional constraint on the timing and manner of Neptune's migration, which severely depleted the mass of the Kuiper Belt \citep{Malhotra1995,Gomesetal2004,Nesvorny2015}.

The magnitude distribution of the Kuiper Belt has long been modeled as a single slope at large sizes \citep[e.g.][]{Jewittetal1996}.
\citet{Gladmanetal2001} found that the smallest TNOs had a size distribution inconsistent with a single power law.
Later, \citet{Bernsteinetal2004} measured a rollover, proving that a single power law was not adequate to describe the observed Kuiper Belt.
Surveys are now reaching deep enough and detecting enough TNOs that additional structure in the size distribution is required to match observations \citep{FuentesHolman2008,Shankmanetal2013,Fraseretal2014,Alexandersenetal2016,Shankmanetal2016}.

Here we focus our analysis on the scattering TNOs and Centaurs. 
Because they come closer to the Sun than most TNOs, we can observe smaller TNOs within this population than any other in the Kuiper Belt. 
Scattering TNOs and Centaurs are part of the dynamically ``hot'' population. 
TNOs in the dynamically hot population have had their orbits excited to higher inclinations and eccentricities by scattering off Neptune or past/current entanglement with mean-motion resonances \citep{Gladman2005}.
Previous work has demonstrated that the hot population, due to its different collisional and formation history, has a different size distribution than the dynamically cold population of the main classical Kuiper belt \citep{Petitetal2011,Fraseretal2014}.
We specifically exclude those TNOs that are currently resonant from the analysis presented in this manuscript, as they are likely to have experienced a different pathway to dynamical excitation than the scattering TNOs and Centaurs \citep[i.e.][]{Gladmanetal2012}.

\citet{Shankmanetal2016} used scattering TNOs detected in four well-characterized surveys to measure the scattering TNO $H$-distribution to great precision.
In this work, we provide an update for the measurement of the scattering TNO $H$-distribution with the inclusion of the full discovery dataset of the Outer Solar System Origins Survey (OSSOS, \citealt{Bannisteretal2016}; the full dataset is in \citealt{Bannisteretal2018}).
OSSOS has completed its observing, more than tripling the sample of scattering TNOs and Centaurs since the analysis of \citet{Shankmanetal2016}.

The analysis here follows on the work of \citet{Shankmanetal2016} using the same methodologies.
We first discuss the OSSOS survey, summarizing the mechanics of the Survey Simulator, which allows us to forward-bias our model to allow statistical comparison with our observational sample of TNOs.
In Section~\ref{sec:measuring}, we summarize the statistical analysis that we use to find the range of acceptable $H$-magnitude distributions allowed by our observed sample. 
Section~\ref{sec:pop} includes our population measurements, and in Section~\ref{sec:discussion} we discuss how our measurements of the scattering disk fit into the larger context of the Solar System.

\section{Scattering Sample Selection} \label{sec:ossos}

\begin{figure}
\begin{center}
\includegraphics[scale=0.55]{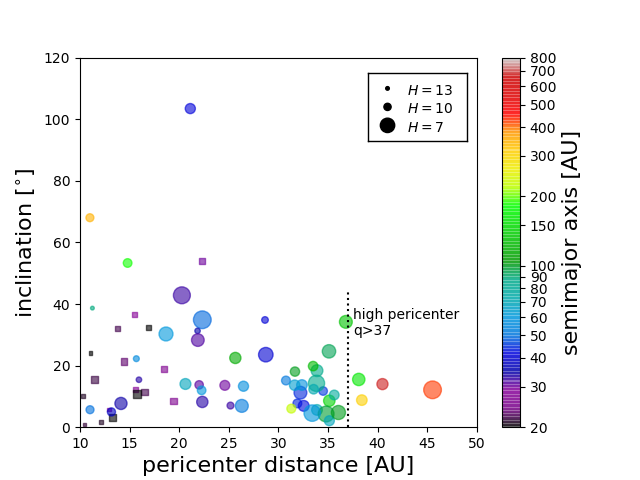}
\caption{
Orbital properties (pericenter distance $q$ and inclination $i$) of the 68 TNOs detected by the OSSOS ensemble of surveys \citep[see Table~\ref{tab:sample} in the Appendix, and Table~3 in][]{Bannisteretal2018} that are classified as scattering (circles) or Centaurs (squares; see Section~\ref{sec:ossos} for details on classification).  
Semimajor axis $a$ is shown via point colour, most of these TNOs have $a<200$~AU.
Point sizes are proportional to diameter (assuming the same albedo);
note that the closest objects are preferentially small due to discovery biases.
Outside $q>37$~AU (noted with dotted line in plot), scattering TNOs have preferentially larger $a$ and are more weakly bound, see text for further discussion.
}
\label{fig:qiplot}
\end{center}
\end{figure}

Because scattering TNOs and Centaurs have high eccentricities, and their pericenter distances can range from nearly Jupiter-crossing to $>$40~AU, the observing biases are extreme and must be accounted for carefully; e.g. small Centaurs and TNOs with closer pericenters are far more likely to be detected in magnitude-limited surveys (as is visible in Figure~\ref{fig:qiplot}).
By using only TNOs detected by well-characterized surveys in this analysis, where the magnitude limits, pointings, and tracking efficiencies are known and published\footnote{Survey Simulator code and OSSOS ensemble survey pointings are publicly available at \url{https://github.com/OSSOS/SurveySimulator}, and properties of TNOs detected by the OSSOS ensemble are published in \citet{Bannisteretal2018}}, we are able to forward-bias models of the scattering disk and statistically compare the resulting biased simulated detections with the real TNO discoveries.

The Outer Solar System Origins Survey (OSSOS) is a large program on the Canada-France Hawaii Telescope over five years to discover TNOs while carefully characterizing tracking fractions, detection efficiencies, and pointing directions, allowing the survey biases to be fully quantified \citep{Bannisteretal2016}.
This methodology has been followed for three other large Kuiper Belt surveys: The Canada-France Ecliptic Plane Survey \citep[CFEPS;][]{Petitetal2011}, the CFEPS high latitude component \citep[HiLat;][]{Petitetal2017}, and the survey of \citet[][hereafter referred to as MA]{Alexandersenetal2016}.  
Combining these three surveys with OSSOS gives a well-characterized set of surveys (which we refer to throughout this paper as the ``OSSOS ensemble''), whose combined detected TNOs provide powerful constraints on the intrinsic TNO orbital distributions and populations when used in combination with the Survey Simulator.  
This statistical reproduction of the survey biases is discussed extensively in other works, \citep[e.g.][]{KavelaarsL3,Petitetal2011,Shankmanetal2016,Bannisteretal2016,Lawleretal2018}.

\citet{Shankmanetal2016} analysed a scattering TNO sample of 22 objects from CFEPS, HiLat, MA, and the first two (of eight) observing blocks of OSSOS. 
OSSOS has since detected dozens of new scattering TNOs and Centaurs, bringing the full sample available for analysis to 68 TNOs (17 Centaurs, 51 scattering).
The orbital elements of the full sample analyzed here are shown in Figure~\ref{fig:qiplot} and Table~\ref{tab:sample} in the Appendix, and further detail is available in Table~3 (ensemble catalogue) in \citet{Bannisteretal2018}.

Here, we use the dynamical classification scheme of \citet{Gladmanetal2008} to determine membership in the scattering and Centaur classes.
These two classes are both unstable on timescales much shorter than the age of the Solar System.
The distinction between them is semimajor axis $a$ relative to Neptune's orbit; Centaurs have smaller $a$ and scattering TNOs have larger (shown by different symbols in Figure~\ref{fig:qiplot}).
Their changes in $a$ over time are usually due to close encounters with one of the giant planets, but can also be due to dynamical diffusion for the more weakly bound TNOs \citep{Bannisteretal2017}: those that have largest pericenter distances $q>37$~AU tend to also have the largest semimajor axes of the sample (see Figure~\ref{fig:qiplot}).
The Centaurs show similar evolution in semimajor axis and represent the low-$a$ tail of the scattering population \citep[e.g.][]{Gomesetal2008}, thus it is expected that they will share the same $H$-distribution.

\section{Measuring the True $H$ Distribution} \label{sec:measuring}

Previous work has shown that there is a sharp transition in the $H$-distribution of the TNOs, though the form of the transition is unclear \citep{Shankmanetal2013,Shankmanetal2016,Fraseretal2014,Alexandersenetal2016}.
We parameterize this transition using a bright-end slope $\alpha_b$, a faint-end slope $\alpha_f$, a break magnitude $H_b$, and a differential contrast $c$.
We use the terminology that $c=1$ is a knee, $c>1$ is a divot.
We refer the reader to Figure~9 in \citet{Shankmanetal2016} for a graphical demonstration of the effect of these two different transitions on the cumulative and differential number distributions in $H$. 

\begin{figure*}[!ht]
\begin{center}
\includegraphics[scale=0.6]{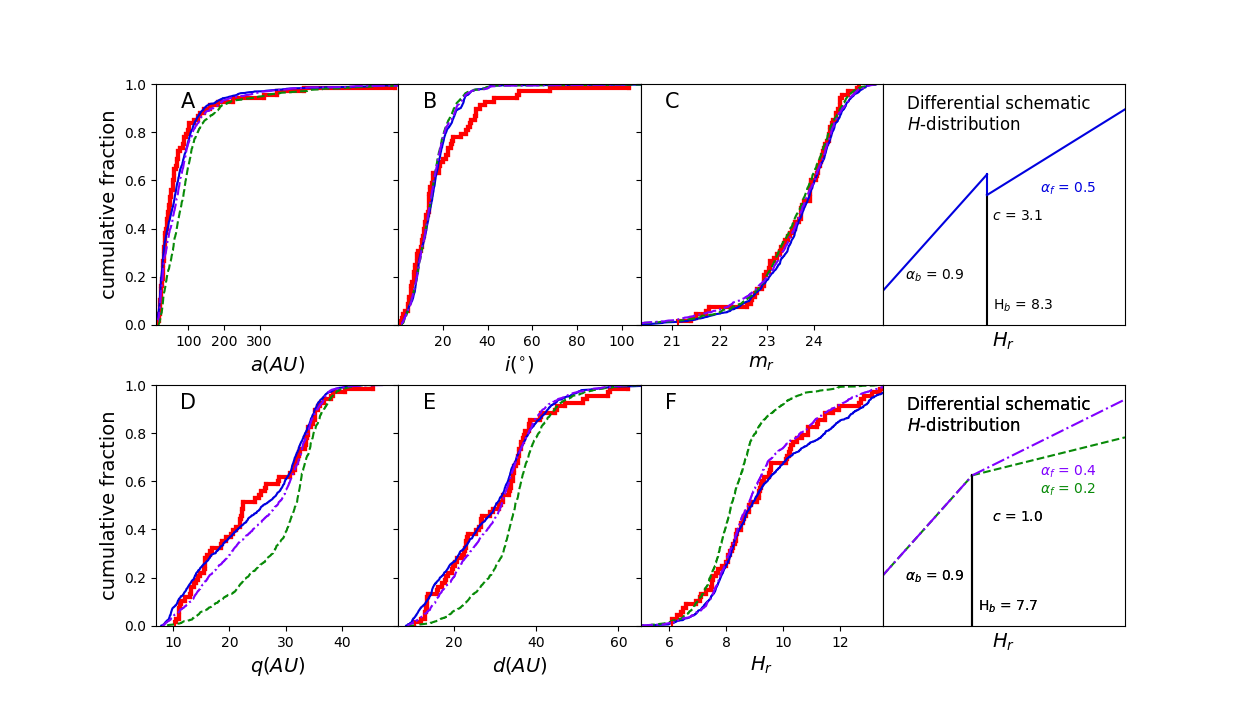}
\caption{
Cumulative distributions across six parameters for the 68 observed scattering TNOs and Centaurs (red step-function), and three candidate $H$-distributions. 
Panels A-F correspond to the semi-major axis $a$, inclination $i$ (see Section~\ref{sec:incl}), magnitude at detection in $r$-band $m_r$, pericenter $q$, distance at detection $d$, and $H$ magnitude in $r$-band, respectively. 
The rightmost panel provides schematics for three different H-distributions: 
(1) our preferred ($c$, $\alpha_f$) pair (solid blue line) 
(2) our preferred knee distribution (dot-dashed purple line) and
(3) the best-fit knee distribution from \citet{Fraseretal2014} (dashed green line).
}
\label{fig:6panel}
\end{center}
\end{figure*}

The slope at the bright end of the TNO $H$-distribution $\alpha_b$, a range of $H\simeq4$--7, is well-probed by previous work \citep[e.g.][]{FraserKavelaars2009,Petitetal2011}.
Our OSSOS ensemble detections range from $H_r$ values of 6 to 14.5 because of the very close pericenter distances of some of these TNOs, and thus this analysis is sensitive to a much fainter $H_r$ range than previous work.  We note that several of the scattering TNOs included in this sample were not observed in $r$-band because some blocks of CFEPS observed only in $g$.  These have had their $g$-band and $H_g$ magnitudes transposed to $r$ by assuming that $g-r=0.7$, which is at the neutral end of the observed color range of dynamically excited TNOs \citep{Tegleretal2016}. \citet{Shankmanetal2016} used $g-r=0.7$, and also demonstrated that using $g-r$ values ranging from 0.5--0.9 makes no difference to the statistical analysis performed below \citep[see Figure~8 in][]{Shankmanetal2016}.

In this analysis, as in previous work \citep{Shankmanetal2013,Shankmanetal2016}, we seek to measure the slope of the faint end of the $H$-distribution $\alpha_f$, the contrast of the transition $c$, and the $H$ magnitude where the break occurs $H_b$. 
We test $H$-distributions from a grid covering $\alpha_f$ from 0.1 to 0.9, and $c$ from 1 to 100, with two different break magnitudes, $H_b=8.3$ \citep[preferred break magnitude from][]{Shankmanetal2016} and $H_b=7.7$ \citep[preferred break magnitude from][]{Fraseretal2014}.

\subsection{The Survey Simulator and Statistical Analysis}

Our method of forward-biasing a model distribution with different $H$-distributions is discussed in detail in \citet{Shankmanetal2013} and \citet{Shankmanetal2016}.  
Briefly, we start with a version of the scattering distribution modelled by the emplacement simulation of \citet{Kaibetal2011}, with the dynamically hotter inclination distribution used in \citet{Shankmanetal2016}.
We then draw orbits from this simulation.
Orbits are randomly oriented (random $\omega$ and $\Omega$), and objects are placed with a random mean anomaly on these orbits (which sets the distance), and are given an $H$-magnitude from within a chosen $H$-distribution, and then an $r$-magnitude is calculated.  
The Survey Simulator then determines if that $r$-magnitude, rate of motion, and on-sky position was detectable in the OSSOS survey ensemble.  
This process is continued until a large number (hundreds) of simulated detections are created.
The cumulative distributions of simulated detections are then statistically compared with the cumulative distributions of the 68 real Centaurs and scattering TNOs in semimajor axis $a$, inclination $i$, $r$-magnitude $m_r$, pericenter distance $q$, distance at detection $d$, and $H$-magnitude in $r$-band $H_r$.  
These six cumulative distributions are shown in Figure~\ref{fig:6panel} for the real TNOs as well as simulated detections using three different $H$-distributions.

\begin{figure*}
\begin{center}
\includegraphics[scale=0.5]{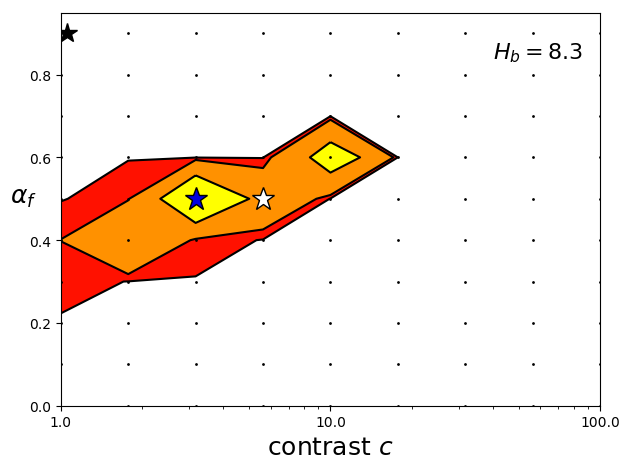} \includegraphics[scale=0.5]{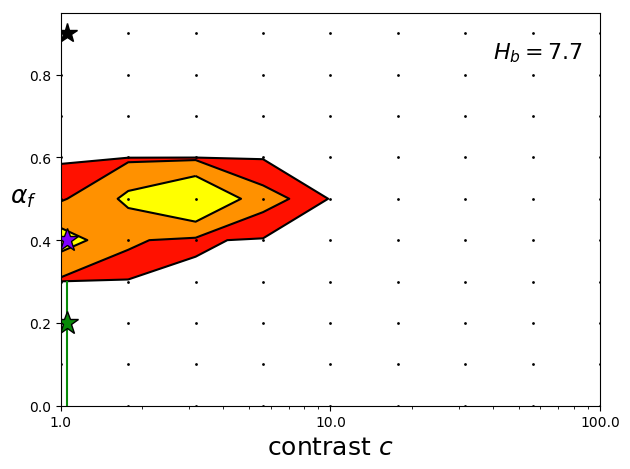}
\caption{
Contours of the rejectability for the tested faint-end slope $\alpha_f$ and contrast $c$ pairs with a break located at $H_b=8.3$ (left) and $H_b=7.7$ (right); all models tested use $\alpha_b=0.9$.
The contours represent the $1\sigma$, $2\sigma$ and $3\sigma$ rejectability levels with white being rejectable at $>3\sigma$, red being rejectable at $>2\sigma$, and orange and yellow not statistically rejectable. 
Stars highlight notable ($\alpha_f$, $c$) pairs:
the dark blue, green, and purple stars show models that are also plotted in the same colour in Figure~\ref{fig:6panel}.
The dark blue star denotes our preferred ($\alpha_f$, $c$) pair (see Section~\ref{sec:best}), the green star (with 1$\sigma$ error bars) denotes the best-fit knee model for dynamically hot TNOs from \citet{Fraseretal2014}, and the purple star is our preferred knee model.
For comparison with previous work, the white star denotes the preferred ($\alpha_f$, $c$) pair from \citet{Shankmanetal2016}, and the black star denotes a single slope of $\alpha=0.9$ (identical in both plots), and is strongly rejectable.
}
\label{fig:contours}
\end{center}
\end{figure*}

The statistical analysis is described in detail in \citet{Shankmanetal2016}, and we summarize below.
We first calculate the Anderson-Darling (AD) statistic \citep{AndersonDarling}, comparing the observed TNOs and the simulated detections for a given $H$-distribution.  
An AD statistic is computed for each parameter.
From previous work, we found that the most powerful lever arms for this analysis (because they vary most for different modelled $H$-distributions) come from using the parameters $q$, $d$, and $H_r$, so we sum the AD statistics calculated for each of these three distributions \citep[following the analysis method of][]{Parker2015}.
This summed AD statistic is bootstrapped by selecting at random 68 objects from the distribution of simulated detections, calculating the AD statistic between this random sample and the simulated detections in each parameter, and summing them. 
This random selection and AD statistic calculation is repeated hundreds of times.
The distribution of summed AD statistics for random samples of the simulated distribution is then compared to the summed AD statistic for the real TNOs.
If that AD statistic or larger occurs for $<5$\% of the random distributions, we can reject that distribution with $>2\sigma$ ($>$95\%) confidence.  To explain in another way, if $<$95\% of random subsets of the model are farther from the parent model than the observations are, then the model cannot be rejected.

\subsubsection{The Scattering Inclination Distribution} \label{sec:incl}

Figure~\ref{fig:6panel} shows a good match between the observations and the preferred model for five of the six parameters measured; the inclination distribution has a rather poor fit at high inclinations (this is true for all $H$-distributions tested).
The paucity of high inclination objects in the model as compared with observations was noted and discussed in \citet{Shankmanetal2016}.  
The difficulty of generating high inclination objects in emplacement models is a well-noted problem \citep[e.g.][]{Kaibetal2011}, and suggests that a small fraction of scattering TNOs may require a different emplacement pathway in order to match the real Kuiper belt.
Suggested mechanisms in the literature include diffusion from the Oort Cloud \citep{Kaibetal2009,Brasseretal2012}, interaction with a distant massive planet \citep{Gomesetal2015}, and interaction with a rouge planet that was later ejected from the Solar System \citep{GladmanChan2006}.
Creating dynamical emplacement models of the Kuiper belt that obtain a realistic inclination distribution is currently an area of active research.  

We perform a simple experiment to make sure that the inclination distribution does not severely affect the three parameters we test ($H$, $d$, and $q$) by doubling and halving all of the inclinations in the model and re-running our statistical test.  
We find that the bootstrapped AD values only vary by 1-2\% for these two very different inclination distributions, and so we conclude that while the inclination distribution shown in Figure~\ref{fig:6panel} does not provide a great match to observations, the other properties of the model still provide an excellent fit to the real scattering TNOs.

\subsection{Preferred $H$-distribution} \label{sec:best}

Using the 68 detected scattering TNOs and Centaurs from the OSSOS ensemble, we find that the least rejectable $H$-distribution is for $\alpha_f=0.5$ and $c=3.2$, using $\alpha_b=0.9$.
This $H$-distribution is shown as a blue solid line in Figure~\ref{fig:6panel}, and by a blue star in Figure~\ref{fig:contours}.

We are unable to statistically reject a knee distribution.  
A transition to a faint slope $\alpha_f=0.4$ at $H_b=7.7$ is non-rejectable at 3$\sigma$ significance in our analysis; this preferred knee $H$-distribution is shown by a purple dash-dotted line in Figure~\ref{fig:6panel}, and by a purple star in Figure~\ref{fig:contours}.
For comparison, the best-fit knee distribution from \citet{Fraseretal2014} is shown with a green star, including 1$\sigma$ error bars.

The preferred divot $H$-distribution from \citet{Shankmanetal2016} remains a viable explanation for the scattering TNO H-distribution (white star in Figure~\ref{fig:contours}), but the analysis here increases the number of rejectable models, more tightly constraining the acceptable parameter space of $\alpha_f$ and $c$.
As in \citet{Shankmanetal2013,Shankmanetal2016}, a single power law ($c=1$, $\alpha_f=\alpha_b=0.9$) is rejectable at $>3\sigma$ significance (shown with a black star in both plots in Figure~\ref{fig:contours}).

Interestingly, we are not able to rule out either break magnitude $H_{b}$ we tested.  
We tested two different values of $H_{b}$: 8.3 and 7.7, based on predictions from previous work \citep{Shankmanetal2016,Fraseretal2014}.
The yellow contours in Figure~\ref{fig:contours} highlight the $H$-distributions which are rejectable by our analysis at the lowest significance (i.e.\ least rejectable distributions).
Contours of $<1\sigma$ rejectability occur for both $H_b$ values that we tested, and ($\alpha_f=0.5$, $c=3.2$) are the least rejectable $H$ distributions for both values of $H_b$.

\section{The Intrinsic Population Size} \label{sec:pop}

We use the Survey Simulator to determine the number of scattering TNOs and Centaurs brighter than a given $H$ magnitude that must be drawn from the \citet{Kaibetal2011} scattering TNO model to allow 68 detections (Table~\ref{tab:pops}: scattering TNOs), or 17 detections from the $a<30$~AU subset of the \citet{Kaibetal2011} scattering TNO model (Table~\ref{tab:popcen}: Centaurs).
Error bars on these intrinsic populations are calculated by running this experiment many times and finding the populations which bracket 95\% of the estimates, the error bars given are thus 95\% confidence intervals on the intrinsic population.

\begin{deluxetable*}{ccccccl}
\tabletypesize{\footnotesize}
\tablecolumns{7}
\tablecaption{Population Estimates for Scattering TNOs \label{tab:pops}}
\tablehead{ & & & $H_r<8.66$ & $H_r<10$ & $H_r<12$ \\
$H_b$ & $\alpha_f$ & $c$ & population & population & population & comment }
\startdata
8.3 & 0.5 & 3.2 & $\left(0.9\pm0.2\right)\times10^5$ & $\left(2.9\pm0.7\right)\times10^5$ & $\left(2.7\pm0.7\right)\times10^6$ & preferred divot, this work \\
7.7 & 0.4 & 1 & $\left(0.8\pm0.2\right)\times10^5$ & $\left(3.5^{+0.9}_{-0.6}\right)\times10^5$ & $\left(2.4^{+0.6}_{-0.4}\right)\times10^6$ & preferred knee, this work \\
8.3 & 0.5 & 5.6 & $\left(1.0^{+0.3}_{-0.2}\right)\times10^5$ & $\left(2.6^{+0.7}_{-0.5}\right)\times10^5$ & $\left(2.1^{+0.6}_{-0.4}\right)\times10^6$ & preferred, \citet{Shankmanetal2016} \\
8.3 & 0.4 & 1 & $\left(0.8\pm0.2\right)\times10^5$ & $\left(4.0\pm0.9\right)\times10^5$ & $\left(2.8^{+0.6}_{-0.7}\right)\times10^6$ & least-rejectable knee, $H_b=8.3$ \\
7.7 & 0.5 & 3.2 & $\left(0.7\pm0.2\right)\times10^5$ & $\left(2.8^{+0.7}_{-0.6}\right)\times10^5$ & $\left(2.7^{+0.7}_{-0.6}\right)\times10^6$ & least-rejectable divot, $H_b=7.7$ \\ \hline
\multicolumn{7}{c}{Previously published population estimates} \\ \hline
8.3 & 0.5 & 5.6 & & & $\sim1\times10^6$ & estimate from \citet{Shankmanetal2013} \\
8.3 & 0.5 & 5.6 & & & $(2.4-8.3)\times10^5$ & estimate from \citet{Shankmanetal2016} \\
$-$ & 0.8 & $-$ & $\left(5^{+5}_{-3}\right)\times10^3$ & & $\left(4^{+4}_{-3}\right)\times10^6$ & CFEPS estimate \citep{Petitetal2011} \\
\enddata
\tablecomments{Error bars on population estimates are 95\% confidence intervals.}
\end{deluxetable*}

\begin{deluxetable*}{ccccccl}
\tabletypesize{\footnotesize}
\tablecolumns{7}
\tablecaption{Population Estimates for Centaurs \label{tab:popcen}}
\tablehead{ & & & $H_r<8.66$ & $H_r<10$ & $H_r<12$ \\
$H_b$ & $\alpha_f$ & $c$ & population & population & population & comment }
\startdata
8.3 & 0.5 & 3.2 & $110^{+60}_{-40}$ & $390^{+200}_{-150}$ & $3500^{+1800}_{-1400} $ & preferred divot, this work \\
7.7 & 0.4 & 1 & $130^{+80}_{-70}$ & $550^{+340}_{-290}$ & $3700^{+2300}_{-2000}$ & preferred knee, this work \\ \hline
\multicolumn{7}{c}{Previously published population estimates} \\ \hline
7.7 & 0.2 & 1 &  $\le75,000$ & $\left(2.8^{+10.0}_{-2.5}\right)\times10^4$ & & calculated from Uranian co-orbitals\tablenotemark{a} \\
8.5 & 0.5 & 6 &  $\le75,000$  & $\left(2.8^{+13.0}_{-2.5}\right)\times10^4$ & & calculated from Uranian co-orbitals\tablenotemark{a} \\
7.7 & 0.2 & 1 & $2500^{+11000}_{-2100}$ & $7100^{+32,000}_{-6800}$ & & calculated from Neptunian co-orbitals\tablenotemark{a} \\
8.5 & 0.5 & 6 & $2900^{+11000}_{-2500}$  & $7500^{+32,000}_{-7100}$ & & calculated from Neptunian co-orbitals\tablenotemark{a} \\
\enddata
\tablecomments{Error bars on population estimates are 95\% confidence intervals.}
\tablenotetext{a}{Calculated from observations and models of \citet{Alexandersenetal2013,Alexandersenetal2016}. Note that populations here are actually for the $a<34$~AU scattering population, a large fraction of which will be Centaurs; see text.}
\end{deluxetable*}

\subsection{The Size of the Scattering TNO Population} \label{sec:scatpop}

Using our preferred $H$-distribution ($H_b=8.3$, $\alpha_f=0.5$, $c=3.2$), the population must be $\left(2.7^{+0.6}_{-0.5}\right)\times10^6$ for $H_r<12$ (which corresponds to $D\gtrsim20$~km for an albedo of 0.04), and $\left(8\pm2\right)\times10^4$ for $H_r<8.66$ (which corresponds to $D\gtrsim100$~km for an albedo of 0.04).  
Interestingly, using other statistically acceptable $H$-distributions does not cause the population to vary by more than a very small factor; the population estimates from all statistically acceptable $H$-distributions are consistent within the 95\% error bars.

Table~\ref{tab:pops} lists population estimates using several different $H$-distributions that are statistically acceptable in our analysis, as well as comparisons with previously published scattering population estimates.
Our population estimates here are slightly higher than those reported in \citet{Shankmanetal2013} and \citet{Shankmanetal2016}.
The Canada-France Ecliptic Plane Survey \citep{Petitetal2011} estimates a population of $5000^{+5000}_{-3000}$ scattering TNOs for $H_g<9.16$ ($H_r\lesssim8.66$), much smaller than our population estimate.  
However, after scaling by the assumed single slope of $\alpha=0.8$ down to $H_r<12$ gives $\left(4^{+4}_{-3}\right)\times10^6$, consistent with our population estimates, albeit with very large error bars.  

Assuming that this size distribution holds for another order of magnitude smaller in TNO size, we can scale our population estimates up to include TNOs at very small sizes ($H<18$), and compare with the number of scattering TNOs that are required to supply the observed population of Jupiter Family Comets (JFCs).
However, this close-in population has been measured to have slightly shallower slopes \citep{Snodgrassetal2011,Fernandezetal2013,Baueretal2017} than the faint slope $\alpha_f$ found in this analysis, so this may not be a valid assumption.
With our preferred $H$-distribution, we find that the scattering population down to $H<18$ should include $3\times10^9$ objects, which is a large enough supply to be the origin of the Jupiter Family Comets \citep{VolkMalhotra2008}.

\subsection{The Size of the Centaur Population} \label{sec:cenpop}

The intrinsic Centaur population is about two orders of magnitude smaller than the intrinsic scattering TNO population, consistent with their shorter dynamical lifetime.
In Table~\ref{tab:popcen}, we compare our Centaur population estimates with the population estimates of temporary Uranian and Neptunian co-orbitals in \citet{Alexandersenetal2016} and the abundance of these relative to $a<34$~AU scattering objects estimated in \citet{Alexandersenetal2013}.
\citet{Alexandersenetal2013} gives the fraction of the $a<34$~AU scattering population that must be trapped as temporary co-orbitals with Neptune and Uranus at any given time.  
The orbital distributions from \citet{Parker2015} and \citet{Alexandersenetal2013} are combined with a knee $H$-distribution similar to the best-fit of \citet{Fraseretal2014} and a divot distribution similar to the preferred $H$-distribution from \citet{Shankmanetal2016} to calculate the population estimates in Table~\ref{tab:popcen}.
The Centaur population estimates from our analysis are much smaller, but are not inconsistent when taking into account the (very large) error bars and upper limits from \citet{Alexandersenetal2013,Alexandersenetal2016}.

Another way we can make use of the Survey Simulator is to estimate how many relatively large Centaurs should exist based on our preferred $H$-distribution.
Using this methodology, we find that the expected number of $H_r<6$ Centaurs is $\leq$1 with 95\% confidence.  
Reassuringly, the largest known Centaur, (10199) Chariklo, has an $H_r$ magnitude of $6.82\pm0.02$ \citep[assuming a linear spectrum and no phase correction;][]{Peixinhoetal2015}.

\section{Discussion} \label{sec:discussion}

Although we are unable to formally reject either a knee or divot distribution, the power of forward-biasing combined with statistical analysis of the full OSSOS dataset has vastly reduced the allowed parameter space compared to previous analyses \citep{Shankmanetal2013,Shankmanetal2016}.
But even with the earlier much smaller number of detections, this analysis technique is powerful.
While the range of parameter space that was non-rejectable in \citet{Shankmanetal2013} was many times larger than in our analysis here, the preferred divot from the analysis in \citet{Shankmanetal2013} still provides a good agreement to the fit obtained here, even though that analysis only included 11 TNOs, while the analysis here contains over six times as many TNOs.

\subsection{Knee or Divot?}

This analysis has shown that a divot fits the data slightly better than a knee distribution, but knees cannot be rejected for several values of $\alpha_f$.
For the break at larger TNO sizes $H_b=7.7$, $\alpha_f$ of 0.4-0.5 and $c$ from 1 (knee) to 5.6 are non-rejectable.
For the $H_b=8.3$ break, a slightly larger parameter space is non-rejectable, encompassing $\alpha_f$ from 0.3 to 0.6, and $c$ from 1 to 10.

Our preferred knee distribution has a slightly steeper slope ($\alpha_f=0.4$) than the best-fit knee $H$-distribution found by the analysis of \citet{Fraseretal2014} ($\alpha_f=0.2$). 
However, the 1$\sigma$ uncertainties the published uncertainties on the \citet{Fraseretal2014} faint-end slope fit allow up to $\alpha_f=0.3$, which is just inside the contour of non-rejectability (Figure~\ref{fig:contours}, right panel), and is thus non-rejectable by our analysis.

\subsection{Comparison with Other TNO Populations}

This analysis is in broad agreement with the luminosity functions found for other dynamically hot populations in the Kuiper belt.

\citet{Fraseretal2014} reports a $\alpha_f$ slope of 0.36, with a break magnitude $H_b=8.4$ for the Trojan asteroids, which is acceptable in our analysis and would thus allow a common $H$-distribution for the two populations.
If the Kuiper belt was emplaced by scattering off the giant planets during a period of instability \citep{Morbidellietal2005,Nesvornyetal2013}, the Trojans would also be drawn from this population and should have the same size distribution \citep{Morbidellietal2009}.  
Determining whether or not the Trojans and dynamically hot Kuiper Belt populations share a size distribution is an important test of this model, and is an area of active research \citep[e.g.][]{WongBrown2015,YoshidaTerai2017}.

While the number of detected Neptune Trojans is small, previous surveys have noted that there appears to be a lack of small members \citep{SheppardTrujillo2010ApJ,Parker2015}, which would be consistent with a divot in the size distribution.

The plutinos (TNOs in the 3:2 mean motion resonance with Neptune) constitute the closest well-populated resonance, so studies are able to probe the size distribution down to smaller sizes than any other resonance.  
The well-characterized survey of \citet{Alexandersenetal2016} performs a similar analysis to this work, and found that a break is required in the plutino size distribution, with a range of contrasts (including 1), break magnitudes, and faint-end slopes that match their $H$-distribution of plutino detections.  
Their preferred divot $H$-distribution is similar to the preferred divot of this work with a steeper faint-end slope: $c=6$ and $\alpha_f=0.8$ at $H_b=8.4$ \citep[though their non-rejectable parameter space covers a large range of $\alpha_f$ and $c$ values, see Figure 10 in][]{Alexandersenetal2016}.
Knee distributions also provide a statistically acceptable match to their plutino detections, with their preferred fit exactly matching ours ($\alpha_f=0.4$ at $H_b=7.7$) and consistent with the best-fit in \citet{Fraseretal2014}.

\citet{Volketal2016}, which used detections only from the first two (of eight) OSSOS observing blocks, find no evidence in favour of a break in the size distribution, but show that this could be an effect of small number statistics.
The analysis of the plutinos in the full OSSOS survey has several times more detected plutinos, and a transition is required in the $H$-distribution to match these observations, however, both a knee or divot transition provide reasonable matches to the data \citep{VolkDPS2017}.

We note that previous analysis of the dynamically hot classical TNOs prefers a bright-end slope $\alpha_b=0.8$ \citep[95\% confidence range 0.6-1.1, see Figure 5 in][]{Petitetal2011}, and ongoing analysis on the OSSOS discoveries indicates perhaps an even shallower slope provides a better fit to the larger TNOs (Petit et al.\ in prep.)
Our bright end slope of $\alpha_b=0.9$ is consistent with our data and with previous analysis of dynamically hot populations \citep[e.g.][]{Gladmanetal2012}, but as more relatively bright TNOs are discovered by current and future all-sky surveys \citep[e.g.,][]{Holmanetal2018}, the best fit for the bright-end slope should be revisited.

\subsection{Comparison with the Cratering Record}

The distribution of craters on a planetary surface can be used to infer the distribution of impactor sizes if one understands the orbital distribution (and thus planetary impact speed distribution) of the projectiles.
Due to its orbital inclination and Kozai oscillation while inside the 3:2 mean-motion, Pluto spends a large fraction of its time at latitudes above the dynamically cold classical belt, and its orbital eccentricity results in it spending little time passing through the cold classical Kuiper belt \citep{Greenstreetetal2015};
the majority of its impacting projectiles thus come from dynamically hot populations, and it is therefore the dynamically hot population's size distribution that will be encoded in the crater counts.

Using imaging from the \emph{New Horizons} spacecraft's Pluto encounter \citep{Sternetal2015,Mooreetal2016} and crater-rate production calculations \citep{Greenstreetetal2015}, the distribution of impactor sizes has been estimated to arise from an $H$-distribution with $\alpha\simeq0.4$ for projectiles with $H$=13--19 \citep[projectile diameters of 1--20~km;][]{Singeretal2016}.
This $H$ range just barely overlaps with our present analysis, that covers $H\simeq$6--13, but this joint data set implies a roughly constant index $\alpha$ could extend from the break near $D\sim$100~km down to $H\simeq19$ ($D\simeq 1$~km).
If the faint-end slope we measure does indeed continue to $D\simeq 1$~km, this is additional support for the scattering disk being the sole source of the JFCs, as this 
assumption was made above (Section~\ref{sec:pop}) to calculate the population size 
that was in agreement with this requirement.
For even smaller objects, recent results of the Charon crater-field analysis indicate that the Kuiper Belt's $\alpha$ becomes even shallower (Singer et al.\ 2018, Science, in review), but sub-km TNOs are beyond the reach of ground-based and even space-based near-Earth telescopes.

\pagebreak
\section{Summary and Conclusions}

This work is an exploration of the scattering TNO $H$-distribution with the full OSSOS sample, expanding on the analysis of \citet{Shankmanetal2016} with a threefold larger set of detections (68 rather than 22 TNOs), and including fainter $H_r$ magnitudes than previous work. 
We have demonstrated that existing models ($H$-distributions with either a divot or knee transition from bright- to faint-end slopes) provide acceptable matches for the $H$-distribution observed for scattering TNOs, but we have greatly constrained the allowed parameter space of possible faint slopes $\alpha_f$ and contrasts of the transition.
Our preferred $H$-distribution has a bright end slope $\alpha_b=0.9$, a faint slope $\alpha_f=0.5$, and a divot of contrast $c=3.2$, though a knee distribution with $\alpha_f=0.4$ is also acceptable.
The $H$ magnitude at the break is not important to our fit, we find equally statistically acceptable $H$-distributions for $H_b=7.7$ or 8.3, both of which were proposed by previous analyses.
Large surveys such as Pan-STARRS and LSST will detect many new TNOs, especially at relatively bright $H$-magnitudes, and that will likely provide more statistical constraint on exactly where the break magnitude is, providing more information on the initial planetesimal formation size and collisional history of the Kuiper Belt.

We find that the shallower slope at faint magnitudes makes populations that are consistent with both the cratering record on Pluto and the population required to be the source of the Jupiter Family Comets. 

A full exploration of possible size distributions would be best done in the context of a formation and evolutionary model of the Solar System. 
The current degeneracy across potential break locations and divot or knee distributions may be addressed through additional constraints from formation theories. 
In order to explore this, one must understand the conditions under which accretion takes place, e.g. born big \citep{Morbidellietal2009bornbig} or pebble accretion \citep{Shannonetal2016}, and must also understand the dynamical excitation process, e.g. whether Neptune's migration was smooth \citep{HahnMalhotra2005}, grainy \citep{NesvornyVokrouhlicky2016}, or chaotic \citep{Tsiganisetal2005}.
By using these dynamical constraints, we can understand the process that emplaced the hot TNOs and shut off collisional grinding, leaving the Kuiper belt with the size distribution we observe today.

\acknowledgments

The authors acknowledge the sacred nature of Maunakea, and appreciate the opportunity to observe from the mountain.
CFHT is operated by the National Research Council (NRC) of Canada, the Institute National des Sciences de l'Universe of the Centre National de la Recherche Scientifique (CNRS) of France, and the University of Hawaii, with OSSOS receiving additional access due to contributions from the Institute of Astronomy and Astrophysics, Academia Sinica, Taiwan.
Data are produced and hosted at the Canadian Astronomy Data Centre; processing and analysis were performed using computing and storage capacity provided by the Canadian Advanced Network For Astronomy Research (CANFAR).

SML gratefully acknowledges support from the NRC-Canada Plaskett Fellowship.  
MTB appreciates support from UK STFC grant ST/L000709/1.
WF acknowledges support from Science and Technology Facilities Council grant ST/P0003094/1.
This project was funded by the National Science and Engineering Research Council and the National Research Council of Canada.

\facility{CFHT (MegaPrime)}
\software{
matplotlib \citep{Hunter2007}, 
scipy \citep{Jonesetal2001}.}

\appendix

The Appendix comprises Table~\ref{tab:sample}.

\begin{deluxetable*}{llccccccl}
\tabletypesize{\footnotesize}
\tablecolumns{9}
\tablecaption{Centaurs and Scattering TNOs in the OSSOS Survey Ensemble\label{tab:sample}}
\tablehead{dynamical	&	Survey	&	$r$-band	&	$H_r$\tablenotemark{b}	&	distance at	&	$a$	&	$e$	&	$i$	&	MPC \\
class\tablenotemark{a} &	name	&	magnitude\tablenotemark{b}	&		&	discovery [AU]	&	[AU]	&		&	[degrees]	&	desig.\tablenotemark{c}}
\startdata
sca	&	o3e01	&	21.50	&	7.73	&	23.291	&	34.416173	&	0.589571	&	7.711	&	K02GG6G	\\
sca	&	o3e11	&	23.60	&	7.86	&	36.851	&	86.729341	&	0.609269	&	18.362	&	K13GD6Z	\\
sca	&	o3l01	&	23.06	&	10.89	&	16.046	&	55.817595	&	0.719066	&	22.246	&	K13U15R	\\
sca	&	o3l65	&	24.14	&	7.51	&	45.138	&	44.608588	&	0.277799	&	11.207	&	K13U16Z	\\
sca	&	o3o14	&	23.54	&	8.00	&	35.456	&	143.317456	&	0.754854	&	8.580	&	K13J64O	\\
sca	&	o3o16	&	23.92	&	8.34	&	35.680	&	57.383825	&	0.435939	&	13.701	&	K13J64P	\\
sca	&	o3o17	&	24.31	&	8.71	&	35.811	&	77.572262	&	0.540647	&	10.459	&	K13J64R	\\
sca	&	o3o36	&	23.73	&	6.09	&	57.342	&	49.020848	&	0.544507	&	34.879	&	K13J64Q	\\
sca	&	o4h03	&	22.69	&	9.55	&	20.758	&	49.901041	&	0.779420	&	5.679	&	K14UM9Q	\\
sca	&	o4h04	&	24.59	&	11.23	&	21.916	&	35.028185	&	0.376088	&	31.276	&	K14UM9A	\\
sca	&	o4h67PD	&	23.07	&	9.49	&	22.886	&	38.083254	&	0.654259	&	4.960	&	K06QI0P	\\
sca	&	o5c002	&	23.74	&	11.18	&	17.958	&	33.555021	&	0.524814	&	15.414	&	$-$	\\
sca	&	o5c022	&	23.68	&	8.30	&	34.284	&	71.897316	&	0.528607	&	5.612	&	$-$	\\
sca	&	o5c101	&	23.79	&	6.58	&	52.291	&	98.388020	&	0.646235	&	4.287	&	$-$	\\
sca	&	o5d002	&	24.95	&	10.36	&	28.844	&	41.040976	&	0.301831	&	34.818	&	$-$	\\
sca	&	o5d020	&	24.54	&	9.14	&	34.655	&	44.202910	&	0.278006	&	7.719	&	$-$	\\
sca	&	o5d025	&	24.19	&	8.60	&	36.217	&	68.621838	&	0.487903	&	2.105	&	$-$	\\
sca	&	o5d034	&	23.91	&	8.08	&	38.181	&	115.493325	&	0.777844	&	22.481	&	$-$	\\
sca	&	o5m03	&	23.94	&	12.85	&	12.879	&	89.174138	&	0.873805	&	38.666	&	$-$	\\
sca	&	o5m04	&	24.38	&	10.19	&	26.018	&	32.488890	&	0.225414	&	7.026	&	$-$	\\
sca	&	o5m52	&	24.27	&	8.12	&	41.057	&	680.202784	&	0.940468	&	13.994	&	K15KG3G	\\
sca	&	o5p009	&	24.07	&	9.20	&	30.845	&	184.132849	&	0.919622	&	53.315	&	$-$	\\
sca	&	o5p019	&	22.94	&	7.55	&	34.605	&	31.378013	&	0.302694	&	28.288	&	$-$	\\
sca	&	o5p021	&	24.71	&	9.27	&	35.180	&	45.967151	&	0.249269	&	11.745	&	$-$	\\
sca	&	o5p024	&	22.80	&	7.30	&	35.900	&	94.674918	&	0.629258	&	24.631	&	$-$	\\
sca	&	o5p025	&	22.66	&	7.08	&	36.250	&	100.871870	&	0.642654	&	4.771	&	$-$	\\
sca	&	o5p060	&	24.46	&	8.34	&	40.983	&	311.768577	&	0.876807	&	8.795	&	K15G50T	\\
sca	&	o5p146	&	24.09	&	6.47	&	57.872	&	85.613291	&	0.604676	&	14.247	&	$-$	\\
sca	&	o5s06	&	22.90	&	8.53	&	26.576	&	56.481339	&	0.531203	&	13.304	&	K15RO5W	\\
sca	&	o5s10	&	24.22	&	8.89	&	33.472	&	101.338298	&	0.687477	&	18.054	&	$-$	\\
sca	&	o5s11	&	24.54	&	9.14	&	33.969	&	50.814125	&	0.394573	&	15.159	&	$-$	\\
sca	&	o5s13	&	24.55	&	9.09	&	34.254	&	226.592608	&	0.861874	&	6.031	&	K15RO5Y	\\
sca	&	o5s20	&	24.04	&	8.24	&	37.139	&	42.894088	&	0.241075	&	6.932	&	$-$	\\
sca	&	o5t04	&	22.99	&	9.32	&	22.722	&	30.988803	&	0.289815	&	13.747	&	K15RO5U	\\
\enddata
\tablecomments{All decimal places listed are significant.  Full dataset available in \citet{Bannisteretal2018}.}
\tablenotetext{a}{Scattering TNOs are designated by ``sca,'' Centaurs by ``cen.''  These and all dynamical classifications within OSSOS use the classification scheme from \citet{Gladmanetal2008}.}
\tablenotetext{b}{As noted in Section~\ref{sec:measuring}, all measurements have been transposed to $r$-band assuming $g-r=0.7$.}
\tablenotetext{c}{\url{https://www.minorplanetcenter.net/iau/info/PackedDes.html}}
\end{deluxetable*}

\setcounter{table}{2}
\begin{deluxetable*}{llccccccl}
\tabletypesize{\footnotesize}
\tablecolumns{9}
\tablecaption{Centaurs and Scattering TNOs in the OSSOS Survey Ensemble, continued \label{tab:sample2}}
\tablehead{dynamical	&	Survey	&	$r$-band	&	$H_r$\tablenotemark{b}	&	distance at	&	$a$	&	$e$	&	$i$	&	MPC \\
class\tablenotemark{a} &	name	&	magnitude\tablenotemark{b}	&		&	discovery [AU]	&	[AU]	&		&	[degrees]	&	desig.\tablenotemark{c}}
\startdata
sca	&	o5t05	&	24.16	&	8.80	&	33.518	&	126.448249	&	0.735055	&	19.83	&	$-$	\\
sca	&	o5t06	&	24.20	&	8.79	&	33.933	&	72.064128	&	0.534457	&	12.327	&	$-$	\\
sca	&	o5t50	&	24.32	&	7.12	&	51.422	&	59.872018	&	0.688095	&	30.267	&	$-$	\\
sca	&	o5t52	&	24.13	&	6.10	&	62.394	&	425.861136	&	0.893065	&	12.138	&	K15RO5X	\\
sca	&	L3h08	&	23.59	&	7.66	&	38.445	&	159.681973	&	0.761413	&	15.500	&	K03H57B	\\
sca	&	L3q01	&	23.30	&	7.46	&	38.171	&	51.054204	&	0.484715	&	6.922	&	K03QB3W	\\
sca	&	L4k09	&	22.94	&	8.63	&	26.634	&	30.191945	&	0.185168	&	13.586	&	K04K18V	\\
sca	&	L4m01	&	23.05	&	8.05	&	31.360	&	33.467236	&	0.332719	&	8.205	&	K04M08W	\\
sca	&	L4p07	&	21.71	&	6.96	&	29.586	&	39.953648	&	0.280856	&	23.545	&	K04PB7Y	\\
sca	&	L4v04	&	23.44	&	8.39	&	31.848	&	64.100391	&	0.506381	&	13.642	&	K04VD1G	\\
sca	&	L4v11	&	23.49	&	9.24	&	26.757	&	60.035908	&	0.629283	&	11.972	&	K04VD1H	\\
sca	&	L4v15	&	21.77	&	8.21	&	22.950	&	68.385618	&	0.698262	&	14.032	&	K04VD1M	\\
sca	&	L7a03	&	23.14	&	6.41	&	46.991	&	59.613266	&	0.439491	&	4.575	&	K06BS4S	\\
sca	&	HL7j2	&	23.37	&	7.50	&	37.377	&	133.932936	&	0.725235	&	34.197	&	K07L38H	\\
sca	&	HL8a1	&	22.93	&	6.29	&	44.517	&	32.392864	&	0.374396	&	42.826	&	K08AD8U	\\
sca	&	HL8n1	&	23.73	&	8.52	&	31.849	&	41.531221	&	0.491379	&	103.447	&	K08K42V	\\
sca	&	HL9m1	&	21.13	&	9.57	&	12.872	&	348.905416	&	0.968470	&	68.016	&	K09M09S	\\
cen	&	o3l02	&	23.91	&	11.47	&	17.045	&	19.327805	&	0.127022	&	32.476	&	K13U17C	\\
cen	&	o3l03	&	24.39	&	10.25	&	25.336	&	25.872108	&	0.249698	&	8.515	&	K13U17U	\\
cen	&	o3o01	&	23.39	&	11.95	&	13.774	&	22.144387	&	0.378570	&	32.021	&	K13J64C	\\
cen	&	o4h01	&	22.74	&	10.29	&	17.756	&	23.195009	&	0.377843	&	21.319	&	K14UM5J	\\
cen	&	o4h02	&	24.33	&	11.47	&	19.526	&	27.954961	&	0.440821	&	12.242	&	K14UM9G	\\
cen	&	o5c001	&	23.72	&	11.75	&	15.857	&	28.529138	&	0.457119	&	36.539	&	$-$	\\
cen	&	o5d001	&	23.93	&	12.74	&	13.286	&	28.271438	&	0.542533	&	5.729	&	$-$	\\
cen	&	o5p001	&	24.05	&	13.40	&	12.029	&	12.048082	&	0.082638	&	24.112	&	$-$	\\
cen	&	o5p003	&	21.39	&	10.15	&	13.563	&	18.145145	&	0.269879	&	3.070	&	$-$	\\
cen	&	o5p004	&	23.92	&	12.68	&	13.563	&	20.995607	&	0.420656	&	1.628	&	$-$	\\
cen	&	o5p005	&	24.34	&	10.67	&	23.501	&	22.225868	&	0.257298	&	11.401	&	$-$	\\
cen	&	o5s04	&	24.51	&	13.11	&	13.441	&	20.915615	&	0.508346	&	10.109	&	$-$	\\
cen	&	o5s05	&	23.21	&	10.10	&	19.884	&	21.981271	&	0.479320	&	15.389	&	K15RO5V	\\
cen	&	o5t02	&	24.91	&	14.51	&	10.616	&	21.692667	&	0.519340	&	0.927	&	$-$	\\
cen	&	o5t03	&	23.27	&	10.48	&	18.515	&	25.967473	&	0.288012	&	18.849	&	$-$	\\
cen	&	mah01	&	24.45	&	10.86	&	22.432	&	30.072429	&	0.259122	&	53.886	&	K12UH7W	\\
cen	&	mal01	&	22.58	&	9.57	&	20.296	&	19.091885	&	0.176854	&	10.811	&	K11Q99F	\\
\enddata
\tablecomments{All decimal places listed are significant.  Full dataset available in \citet{Bannisteretal2018}.}
\tablenotetext{a}{Scattering TNOs are designated by ``sca,'' Centaurs by ``cen.''  These and all dynamical classifications within OSSOS use the classification scheme from \citet{Gladmanetal2008}.}
\tablenotetext{b}{As noted in Section~\ref{sec:measuring}, all measurements have been transposed to $r$-band assuming $g-r=0.7$.}
\tablenotetext{c}{\url{https://www.minorplanetcenter.net/iau/info/PackedDes.html}}
\end{deluxetable*}

\bibliographystyle{yahapj}

\begin{thebibliography}{}
\providecommand\natexlab[1]{#1}
\providecommand\JournalTitle[1]{#1}

\bibitem[{{Alexandersen} {et~al.}(2013){Alexandersen}, {Gladman},
  {Greenstreet}, {Kavelaars}, {Petit}, \& {Gwyn}}]{Alexandersenetal2013}
{Alexandersen}, M., {Gladman}, B., {Greenstreet}, S., {et~al.} 2013,
  \href{http://dx.doi.org/10.1126/science.1238072}{\JournalTitle{Science}, 341,
  994}

\bibitem[{{Alexandersen} {et~al.}(2016){Alexandersen}, {Gladman}, {Kavelaars},
  {Petit}, {Gwyn}, {Shankman}, \& {Pike}}]{Alexandersenetal2016}
{Alexandersen}, M., {Gladman}, B., {Kavelaars}, J.~J., {et~al.} 2016,
  \href{http://dx.doi.org/10.3847/0004-6256/152/5/111}{\JournalTitle{\aj}, 152,
  111}

\bibitem[{Anderson \& Darling(1954)}]{AndersonDarling}
Anderson, T.~W., \& Darling, D.~A. 1954,
  \href{http://dx.doi.org/10.1080/01621459.1954.10501232}{\JournalTitle{Journal
  of the American Statistical Association}, 49, 765}

\bibitem[{{Bannister} {et~al.}(2016){Bannister}, {Kavelaars}, {Petit},
  {Gladman}, {Gwyn}, {Chen}, {Volk}, {Alexandersen}, {Benecchi}, {Delsanti},
  {Fraser}, {Granvik}, {Grundy}, {Guilbert-Lepoutre}, {Hestroffer}, {Ip},
  {Jakubik}, {Jones}, {Kaib}, {Kavelaars}, {Lacerda}, {Lawler}, {Lehner},
  {Lin}, {Lister}, {Lykawka}, {Monty}, {Marsset}, {Murray-Clay}, {Noll},
  {Parker}, {Pike}, {Rousselot}, {Rusk}, {Schwamb}, {Shankman}, {Sicardy},
  {Vernazza}, \& {Wang}}]{Bannisteretal2016}
{Bannister}, M.~T., {Kavelaars}, J.~J., {Petit}, J.-M., {et~al.} 2016,
  \href{http://dx.doi.org/10.3847/0004-6256/152/3/70}{\JournalTitle{\aj}, 152,
  70}

\bibitem[{{Bannister} {et~al.}(2017){Bannister}, {Shankman}, {Volk}, {Chen},
  {Kaib}, {Gladman}, {Jakubik}, {Kavelaars}, {Fraser}, {Schwamb}, {Petit},
  {Wang}, {Gwyn}, {Alexandersen}, \& {Pike}}]{Bannisteretal2017}
{Bannister}, M.~T., {Shankman}, C., {Volk}, K., {et~al.} 2017,
  \href{http://dx.doi.org/10.3847/1538-3881/aa6db5}{\JournalTitle{\aj}, 153,
  262}

\bibitem[{{Bannister} {et~al.}(2018){Bannister}, {Gladman}, {Kavelaars},
  {Petit}, {Volk}, {Chen}, {Alexandersen}, {Gwyn}, {Delsanti}, {Fraser},
  {Granvik}, {Jakubik}, {Kaib}, {Lawler}, {Marsset}, {Pike}, {Shankman},
  {Thirouin}, \& {Vernazza}}]{Bannisteretal2018}
{Bannister}, M.~T., {Gladman}, B.~J., {Kavelaars}, J.~J., {et~al.} 2018,
  \href{http://dx.doi.org/10.3847/0004-6256/152/3/70}{\JournalTitle{\aj}, 152,
  70}

\bibitem[{{Bauer} {et~al.}(2017){Bauer}, {Grav}, {Fern{\'a}ndez}, {Mainzer},
  {Kramer}, {Masiero}, {Spahr}, {Nugent}, {Stevenson}, {Meech}, {Cutri},
  {Lisse}, {Walker}, {Dailey}, {Rosser}, {Krings}, {Ruecker}, {Wright}, \& {the
  NEOWISE Team}}]{Baueretal2017}
{Bauer}, J.~M., {Grav}, T., {Fern{\'a}ndez}, Y.~R., {et~al.} 2017,
  \href{http://dx.doi.org/10.3847/1538-3881/aa72df}{\JournalTitle{\aj}, 154,
  53}

\bibitem[{{Bernstein} {et~al.}(2004){Bernstein}, {Trilling}, {Allen}, {Brown},
  {Holman}, \& {Malhotra}}]{Bernsteinetal2004}
{Bernstein}, G.~M., {Trilling}, D.~E., {Allen}, R.~L., {et~al.} 2004,
  \href{http://dx.doi.org/10.1086/422919}{\JournalTitle{\aj}, 128, 1364}

\bibitem[{{Bottke} {et~al.}(2005){Bottke}, {Durda}, {Nesvorn{\'y}}, {Jedicke},
  {Morbidelli}, {Vokrouhlick{\'y}}, \& {Levison}}]{Bottkeetal2005}
{Bottke}, W.~F., {Durda}, D.~D., {Nesvorn{\'y}}, D., {et~al.} 2005,
  \href{http://dx.doi.org/10.1016/j.icarus.2004.10.026}{\JournalTitle{\icarus},
  175, 111}

\bibitem[{{Brasser} {et~al.}(2012){Brasser}, {Schwamb}, {Lykawka}, \&
  {Gomes}}]{Brasseretal2012}
{Brasser}, R., {Schwamb}, M.~E., {Lykawka}, P.~S., \& {Gomes}, R.~S. 2012,
  \href{http://dx.doi.org/10.1111/j.1365-2966.2011.20264.x}{\JournalTitle{\mnras},
  420, 3396}

\bibitem[{{Dohnanyi}(1969)}]{Dohnanyi1969}
{Dohnanyi}, J.~S. 1969,
  \href{http://dx.doi.org/10.1029/JB074i010p02531}{\JournalTitle{\jgr}, 74,
  2531}

\bibitem[{{Duffard} {et~al.}(2014){Duffard}, {Pinilla-Alonso}, {Santos-Sanz},
  {Vilenius}, {Ortiz}, {Mueller}, {Fornasier}, {Lellouch}, {Mommert}, {Pal},
  {Kiss}, {Mueller}, {Stansberry}, {Delsanti}, {Peixinho}, \&
  {Trilling}}]{Duffardetal2014}
{Duffard}, R., {Pinilla-Alonso}, N., {Santos-Sanz}, P., {et~al.} 2014,
  \href{http://dx.doi.org/10.1051/0004-6361/201322377}{\JournalTitle{\aap},
  564, A92}

\bibitem[{{Fern{\'a}ndez} {et~al.}(2013){Fern{\'a}ndez}, {Kelley}, {Lamy},
  {Toth}, {Groussin}, {Lisse}, {A'Hearn}, {Bauer}, {Campins}, {Fitzsimmons},
  {Licandro}, {Lowry}, {Meech}, {Pittichov{\'a}}, {Reach}, {Snodgrass}, \&
  {Weaver}}]{Fernandezetal2013}
{Fern{\'a}ndez}, Y.~R., {Kelley}, M.~S., {Lamy}, P.~L., {et~al.} 2013,
  \href{http://dx.doi.org/10.1016/j.icarus.2013.07.021}{\JournalTitle{\icarus},
  226, 1138}

\bibitem[{{Fraser} {et~al.}(2014){Fraser}, {Brown}, {Morbidelli}, {Parker}, \&
  {Batygin}}]{Fraseretal2014}
{Fraser}, W.~C., {Brown}, M.~E., {Morbidelli}, A., {Parker}, A., \& {Batygin},
  K. 2014,
  \href{http://dx.doi.org/10.1088/0004-637X/782/2/100}{\JournalTitle{\apj},
  782, 100}

\bibitem[{{Fraser} \& {Kavelaars}(2009)}]{FraserKavelaars2009}
{Fraser}, W.~C., \& {Kavelaars}, J.~J. 2009,
  \href{http://dx.doi.org/10.1088/0004-6256/137/1/72}{\JournalTitle{\aj}, 137,
  72}

\bibitem[{{Fraser} {et~al.}(2008){Fraser}, {Kavelaars}, {Holman}, {Pritchet},
  {Gladman}, {Grav}, {Jones}, {MacWilliams}, \& {Petit}}]{Fraseretal2008}
{Fraser}, W.~C., {Kavelaars}, J.~J., {Holman}, M.~J., {et~al.} 2008,
  \href{http://dx.doi.org/10.1016/j.icarus.2008.01.014}{\JournalTitle{\icarus},
  195, 827}

\bibitem[{{Fuentes} \& {Holman}(2008)}]{FuentesHolman2008}
{Fuentes}, C.~I., \& {Holman}, M.~J. 2008,
  \href{http://dx.doi.org/10.1088/0004-6256/136/1/83}{\JournalTitle{\aj}, 136,
  83}

\bibitem[{{Gladman}(2005)}]{Gladman2005}
{Gladman}, B. 2005,
  \href{http://dx.doi.org/10.1126/science.1100553}{\JournalTitle{Science}, 307,
  71}

\bibitem[{{Gladman} \& {Chan}(2006)}]{GladmanChan2006}
{Gladman}, B., \& {Chan}, C. 2006,
  \href{http://dx.doi.org/10.1086/505214}{\JournalTitle{\apjl}, 643, L135}

\bibitem[{{Gladman} {et~al.}(2001){Gladman}, {Kavelaars}, {Petit},
  {Morbidelli}, {Holman}, \& {Loredo}}]{Gladmanetal2001}
{Gladman}, B., {Kavelaars}, J.~J., {Petit}, J.-M., {et~al.} 2001,
  \href{http://dx.doi.org/10.1086/322080}{\JournalTitle{\aj}, 122, 1051}

\bibitem[{{Gladman} {et~al.}(2008){Gladman}, {Marsden}, \&
  {Vanlaerhoven}}]{Gladmanetal2008}
{Gladman}, B., {Marsden}, B.~G., \& {Vanlaerhoven}, C. 2008, {Nomenclature in
  the Outer Solar System}, ed. {Barucci, M.~A., Boehnhardt, H., Cruikshank,
  D.~P., Morbidelli, A., \& Dotson, R.}, 43

\bibitem[{{Gladman} {et~al.}(2012){Gladman}, {Lawler}, {Petit}, {Kavelaars},
  {Jones}, {Parker}, {Van Laerhoven}, {Nicholson}, {Rousselot}, {Bieryla}, \&
  {Ashby}}]{Gladmanetal2012}
{Gladman}, B., {Lawler}, S.~M., {Petit}, J.-M., {et~al.} 2012,
  \href{http://dx.doi.org/10.1088/0004-6256/144/1/23}{\JournalTitle{\aj}, 144,
  23}

\bibitem[{{Gomes} {et~al.}(2004){Gomes}, {Morbidelli}, \&
  {Levison}}]{Gomesetal2004}
{Gomes}, R.~S., {Morbidelli}, A., \& {Levison}, H.~F. 2004,
  \href{http://dx.doi.org/10.1016/j.icarus.2004.03.011}{\JournalTitle{\icarus},
  170, 492}
  
\bibitem[{{Gomes} {et al.}(2008){Gomes}, {Fern{\'a}ndez}, {Gallardo}, \& {Brunini}}]{Gomesetal2008} 
{Gomes}, R.~S., {Fern{\'a}ndez}, J.~A., {Gallardo}, T., \& {Brunini}, A.\ 2008, The Solar System Beyond Neptune, 259

\bibitem[{{Gomes} {et~al.}(2015){Gomes}, {Soares}, \&
  {Brasser}}]{Gomesetal2015}
{Gomes}, R.~S., {Soares}, J.~S., \& {Brasser}, R. 2015,
  \href{http://dx.doi.org/10.1016/j.icarus.2015.06.020}{\JournalTitle{\icarus},
  258, 37}

\bibitem[{{Greenstreet} {et~al.}(2015){Greenstreet}, {Gladman}, \&
  {McKinnon}}]{Greenstreetetal2015}
{Greenstreet}, S., {Gladman}, B., \& {McKinnon}, W.~B. 2015,
  \href{http://dx.doi.org/10.1016/j.icarus.2015.05.026}{\JournalTitle{\icarus},
  258, 267}

\bibitem[{{Hahn} \& {Malhotra}(2005)}]{HahnMalhotra2005}
{Hahn}, J.~M., \& {Malhotra}, R. 2005,
  \href{http://dx.doi.org/10.1086/452638}{\JournalTitle{\aj}, 130, 2392}
  
\bibitem[{{Holman} {et~al.}(2018){Holman}, {Payne}, {Fraser}, {et~al.}}]{Holmanetal2018} 
{Holman}, M.~J., {Payne}, M.~J., {Fraser}, W., et al.\ 2018, \apjl, 855, L6

\bibitem[{Hunter(2007)}]{Hunter2007}
Hunter, J.~D. 2007, \JournalTitle{Computing In Science \& Engineering}, 9, 90

\bibitem[{{Irwin} {et~al.}(1995){Irwin}, {Tremaine}, \&
  {Zytkow}}]{Irwinetal1995}
{Irwin}, M., {Tremaine}, S., \& {Zytkow}, A.~N. 1995,
  \href{http://dx.doi.org/10.1086/117749}{\JournalTitle{\aj}, 110, 3082}

\bibitem[{{Jewitt} {et~al.}(1996){Jewitt}, {Luu}, \& {Chen}}]{Jewittetal1996}
{Jewitt}, D., {Luu}, J., \& {Chen}, J. 1996,
  \href{http://dx.doi.org/10.1086/118093}{\JournalTitle{\aj}, 112, 1225}

\bibitem[{{Jewitt} {et~al.}(2010){Jewitt}, {Weaver}, {Agarwal}, {Mutchler}, \&
  {Drahus}}]{Jewittetal2010}
{Jewitt}, D., {Weaver}, H., {Agarwal}, J., {Mutchler}, M., \& {Drahus}, M.
  2010, \href{http://dx.doi.org/10.1038/nature09456}{\JournalTitle{\nat}, 467,
  817}

\bibitem[{{Jones} {et~al.}(2001){Jones}, {Oliphant}, {Peterson}, \&
  Others}]{Jonesetal2001}
{Jones}, E., {Oliphant}, T., {Peterson}, P., \& Others. 2001, SciPy: Open
  source scientific tools for Python

\bibitem[{{Kaib} {et~al.}(2011){Kaib}, {Ro{\v s}kar}, \&
  {Quinn}}]{Kaibetal2011}
{Kaib}, N.~A., {Ro{\v s}kar}, R., \& {Quinn}, T. 2011,
  \href{http://dx.doi.org/10.1016/j.icarus.2011.07.037}{\JournalTitle{\icarus},
  215, 491}

\bibitem[{{Kaib} {et~al.}(2009){Kaib}, {Becker}, {Jones}, {Puckett}, {Bizyaev},
  {Dilday}, {Frieman}, {Oravetz}, {Pan}, {Quinn}, {Schneider}, \&
  {Watters}}]{Kaibetal2009}
{Kaib}, N.~A., {Becker}, A.~C., {Jones}, R.~L., {et~al.} 2009,
  \href{http://dx.doi.org/10.1088/0004-637X/695/1/268}{\JournalTitle{\apj},
  695, 268}

\bibitem[{{Kavelaars} {et~al.}(2009){Kavelaars}, {Jones}, {Gladman}, {Petit},
  {Parker}, {Van Laerhoven}, {Nicholson}, {Rousselot}, {Scholl}, {Mousis},
  {Marsden}, {Benavidez}, {Bieryla}, {Campo Bagatin}, {Doressoundiram},
  {Margot}, {Murray}, \& {Veillet}}]{KavelaarsL3}
{Kavelaars}, J.~J., {Jones}, R.~L., {Gladman}, B.~J., {et~al.} 2009,
  \href{http://dx.doi.org/10.1088/0004-6256/137/6/4917}{\JournalTitle{\aj},
  137, 4917}
  
 \bibitem[{{Lawler} {et~al.}(2018){Lawler}, {Kavelaars}, {Alexandersen}, 
   {Bannister}, {Gladman}, {Petit}, \& {Shankman}}]{Lawleretal2018} 
   {Lawler}, S.~M., {Kavelaars}, J., {Alexandersen}, M., {et~al.} 2018, arXiv:1802.00460 


\bibitem[{{Malhotra}(1995)}]{Malhotra1995}
{Malhotra}, R. 1995,
  \href{http://dx.doi.org/10.1086/117532}{\JournalTitle{\aj}, 110, 420}

\bibitem[{{Moore} {et~al.}(2016){Moore}, {McKinnon}, {Spencer}, {Howard},
  {Schenk}, {Beyer}, {Nimmo}, {Singer}, {Umurhan}, {White}, {Stern}, {Ennico},
  {Olkin}, {Weaver}, {Young}, {Binzel}, {Buie}, {Buratti}, {Cheng},
  {Cruikshank}, {Grundy}, {Linscott}, {Reitsema}, {Reuter}, {Showalter},
  {Bray}, {Chavez}, {Howett}, {Lauer}, {Lisse}, {Parker}, {Porter}, {Robbins},
  {Runyon}, {Stryk}, {Throop}, {Tsang}, {Verbiscer}, {Zangari}, {Chaikin},
  {Wilhelms}, {Bagenal}, {Gladstone}, {Andert}, {Andrews}, {Banks}, {Bauer},
  {Bauman}, {Barnouin}, {Bedini}, {Beisser}, {Bhaskaran}, {Birath}, {Bird},
  {Bogan}, {Bowman}, {Brozovic}, {Bryan}, {Buckley}, {Bushman}, {Calloway},
  {Carcich}, {Conard}, {Conrad}, {Cook}, {Custodio}, {Ore}, {Deboy},
  {Dischner}, {Dumont}, {Earle}, {Elliott}, {Ercol}, {Ernst}, {Finley},
  {Flanigan}, {Fountain}, {Freeze}, {Greathouse}, {Green}, {Guo}, {Hahn},
  {Hamilton}, {Hamilton}, {Hanley}, {Harch}, {Hart}, {Hersman}, {Hill}, {Hill},
  {Hinson}, {Holdridge}, {Horanyi}, {Jackman}, {Jacobson}, {Jennings},
  {Kammer}, {Kang}, {Kaufmann}, {Kollmann}, {Krimigis}, {Kusnierkiewicz},
  {Lee}, {Lindstrom}, {Lunsford}, {Mallder}, {Martin}, {McComas}, {McNutt},
  {Mehoke}, {Mehoke}, {Melin}, {Mutchler}, {Nelson}, {Nunez}, {Ocampo}, {Owen},
  {Paetzold}, {Page}, {Parker}, {Pelletier}, {Peterson}, {Pinkine}, {Piquette},
  {Protopapa}, {Redfern}, {Roberts}, {Rogers}, {Rose}, {Retherford},
  {Ryschkewitsch}, {Schindhelm}, {Sepan}, {Soluri}, {Stanbridge}, {Steffl},
  {Strobel}, {Summers}, {Szalay}, {Tapley}, {Taylor}, {Taylor}, {Tyler},
  {Versteeg}, {Vincent}, {Webbert}, {Weidner}, {Weigle}, {Whittenburg},
  {Williams}, {Williams}, {Williams}, {Woods}, \& {Zirnstein}}]{Mooreetal2016}
{Moore}, J.~M., {McKinnon}, W.~B., {Spencer}, J.~R., {et~al.} 2016,
  \href{http://dx.doi.org/10.1126/science.aad7055}{\JournalTitle{Science}, 351,
  1284}

\bibitem[{{Morbidelli} {et~al.}(2009{\natexlab{a}}){Morbidelli}, {Bottke},
  {Nesvorn{\'y}}, \& {Levison}}]{Morbidellietal2009bornbig}
{Morbidelli}, A., {Bottke}, W.~F., {Nesvorn{\'y}}, D., \& {Levison}, H.~F.
  2009{\natexlab{a}},
  \href{http://dx.doi.org/10.1016/j.icarus.2009.07.011}{\JournalTitle{\icarus},
  204, 558}

\bibitem[{{Morbidelli} {et~al.}(2009{\natexlab{b}}){Morbidelli}, {Levison},
  {Bottke}, {Dones}, \& {Nesvorn{\'y}}}]{Morbidellietal2009}
{Morbidelli}, A., {Levison}, H.~F., {Bottke}, W.~F., {Dones}, L., \&
  {Nesvorn{\'y}}, D. 2009{\natexlab{b}},
  \href{http://dx.doi.org/10.1016/j.icarus.2009.02.033}{\JournalTitle{\icarus},
  202, 310}

\bibitem[{{Morbidelli} {et~al.}(2005){Morbidelli}, {Levison}, {Tsiganis}, \&
  {Gomes}}]{Morbidellietal2005}
{Morbidelli}, A., {Levison}, H.~F., {Tsiganis}, K., \& {Gomes}, R. 2005,
  \href{http://dx.doi.org/10.1038/nature03540}{\JournalTitle{\nat}, 435, 462}

\bibitem[{{Nesvorn{\'y}}(2015)}]{Nesvorny2015}
{Nesvorn{\'y}}, D. 2015,
  \href{http://dx.doi.org/10.1088/0004-6256/150/3/73}{\JournalTitle{\aj}, 150,
  73}

\bibitem[{{Nesvorn{\'y}} \&
  {Vokrouhlick{\'y}}(2016)}]{NesvornyVokrouhlicky2016}
{Nesvorn{\'y}}, D., \& {Vokrouhlick{\'y}}, D. 2016,
  \href{http://dx.doi.org/10.3847/0004-637X/825/2/94}{\JournalTitle{\apj}, 825,
  94}

\bibitem[{{Nesvorn{\'y}} {et~al.}(2013){Nesvorn{\'y}}, {Vokrouhlick{\'y}}, \&
  {Morbidelli}}]{Nesvornyetal2013}
{Nesvorn{\'y}}, D., {Vokrouhlick{\'y}}, D., \& {Morbidelli}, A. 2013,
  \href{http://dx.doi.org/10.1088/0004-637X/768/1/45}{\JournalTitle{\apj}, 768,
  45}

\bibitem[{{Pan} \& {Schlichting}(2012)}]{PanSchlichting2012}
{Pan}, M., \& {Schlichting}, H.~E. 2012,
  \href{http://dx.doi.org/10.1088/0004-637X/747/2/113}{\JournalTitle{\apj},
  747, 113}

\bibitem[{{Parker}(2015)}]{Parker2015}
{Parker}, A.~H. 2015,
  \href{http://dx.doi.org/10.1016/j.icarus.2014.09.043}{\JournalTitle{\icarus},
  247, 112}
  
  \bibitem[{{Peixinho} {et~al.}(2015){Peixinho}, {Delsanti}, \& {Doressoundiram}}]{Peixinhoetal2015} Peixinho, N., Delsanti, A., \& Doressoundiram, A.\ 2015, \aap, 577, A35

\bibitem[{{Petit} {et~al.}(2011){Petit}, {Kavelaars}, {Gladman}, {Jones},
  {Parker}, {Van Laerhoven}, {Nicholson}, {Mars}, {Rousselot}, {Mousis},
  {Marsden}, {Bieryla}, {Murray}, {Ashby}, {Benavidez}, {Campo Bagatin}, \&
  {Veillet}}]{Petitetal2011}
{Petit}, J., {Kavelaars}, J.~J., {Gladman}, B., {et~al.} 2011,
  \JournalTitle{\aj}

\bibitem[{{Petit} {et~al.}(2008){Petit}, {Kavelaars}, {Gladman}, \&
  {Loredo}}]{Petitetal2008}
{Petit}, J.-M., {Kavelaars}, J.~J., {Gladman}, B., \& {Loredo}, T. 2008, {Size
  Distribution of Multikilometer Transneptunian Objects}, ed. M.~A. {Barucci},
  H.~{Boehnhardt}, D.~P. {Cruikshank}, A.~{Morbidelli}, \& R.~{Dotson}, 71

\bibitem[{{Petit} {et~al.}(2017){Petit}, {Kavelaars}, {Gladman}, {Jones},
  {Parker}, {Bieryla}, {Van Laerhoven}, {Pike}, {Nicholson}, {Ashby}, \&
  {Lawler}}]{Petitetal2017}
{Petit}, J.-M., {Kavelaars}, J.~J., {Gladman}, B.~J., {et~al.} 2017,
  \href{http://dx.doi.org/10.3847/1538-3881/aa6aa5}{\JournalTitle{\aj}, 153,
  236}

\bibitem[{{Schlichting} {et~al.}(2013){Schlichting}, {Fuentes}, \&
  {Trilling}}]{Schlichtingetal2013}
{Schlichting}, H.~E., {Fuentes}, C.~I., \& {Trilling}, D.~E. 2013,
  \href{http://dx.doi.org/10.1088/0004-6256/146/2/36}{\JournalTitle{\aj}, 146,
  36}

\bibitem[{{Shankman} {et~al.}(2013){Shankman}, {Gladman}, {Kaib}, {Kavelaars},
  \& {Petit}}]{Shankmanetal2013}
{Shankman}, C., {Gladman}, B.~J., {Kaib}, N., {Kavelaars}, J.~J., \& {Petit},
  J.~M. 2013,
  \href{http://dx.doi.org/10.1088/2041-8205/764/1/L2}{\JournalTitle{\apjl},
  764, L2}

\bibitem[{{Shankman} {et~al.}(2016){Shankman}, {Kavelaars}, {Gladman},
  {Alexandersen}, {Kaib}, {Petit}, {Bannister}, {Chen}, {Gwyn}, {Jakubik}, \&
  {Volk}}]{Shankmanetal2016}
{Shankman}, C., {Kavelaars}, J., {Gladman}, B.~J., {et~al.} 2016,
  \href{http://dx.doi.org/10.3847/0004-6256/151/2/31}{\JournalTitle{\aj}, 151,
  31}

\bibitem[{{Shannon} {et~al.}(2016){Shannon}, {Wu}, \&
  {Lithwick}}]{Shannonetal2016}
{Shannon}, A., {Wu}, Y., \& {Lithwick}, Y. 2016,
  \href{http://dx.doi.org/10.3847/0004-637X/818/2/175}{\JournalTitle{\apj},
  818, 175}

\bibitem[{{Sheppard} \& {Trujillo}(2010)}]{SheppardTrujillo2010ApJ}
{Sheppard}, S.~S., \& {Trujillo}, C.~A. 2010,
  \href{http://dx.doi.org/10.1088/2041-8205/723/2/L233}{\JournalTitle{\apjl},
  723, L233}

\bibitem[{{Singer} {et~al.}(2016){Singer}, {McKinnon}, {Robbins}, {Schenk},
  {Greenstreet}, {Gladman}, {Parker}, {Stern}, {Bray}, {Weaver}, {Beyer},
  {Young}, {Spencer}, {Moore}, {Olkin}, {Ennico}, {Binzel}, {Grundy}, {New
  Horizons Geology}, {Geophysics Team}, {New Horizons Composition Team}, {New
  Horizons Mvic Team}, \& {New Horizons Lorri Team}}]{Singeretal2016}
{Singer}, K.~N., {McKinnon}, W.~B., {Robbins}, S.~J., {et~al.} 2016, in Lunar
  and Planetary Inst.~Technical Report, Vol.~47, Lunar and Planetary Science
  Conference, 2310

\bibitem[{{Snodgrass} {et~al.}(2011){Snodgrass}, {Fitzsimmons}, {Lowry}, \&
  {Weissman}}]{Snodgrassetal2011}
{Snodgrass}, C., {Fitzsimmons}, A., {Lowry}, S.~C., \& {Weissman}, P. 2011,
  \href{http://dx.doi.org/10.1111/j.1365-2966.2011.18406.x}{\JournalTitle{\mnras},
  414, 458}

\bibitem[{{Stern} {et~al.}(2015){Stern}, {Bagenal}, {Ennico}, {Gladstone},
  {Grundy}, {McKinnon}, {Moore}, {Olkin}, {Spencer}, {Weaver}, {Young},
  {Andert}, {Andrews}, {Banks}, {Bauer}, {Bauman}, {Barnouin}, {Bedini},
  {Beisser}, {Beyer}, {Bhaskaran}, {Binzel}, {Birath}, {Bird}, {Bogan},
  {Bowman}, {Bray}, {Brozovic}, {Bryan}, {Buckley}, {Buie}, {Buratti},
  {Bushman}, {Calloway}, {Carcich}, {Cheng}, {Conard}, {Conrad}, {Cook},
  {Cruikshank}, {Custodio}, {Dalle Ore}, {Deboy}, {Dischner}, {Dumont},
  {Earle}, {Elliott}, {Ercol}, {Ernst}, {Finley}, {Flanigan}, {Fountain},
  {Freeze}, {Greathouse}, {Green}, {Guo}, {Hahn}, {Hamilton}, {Hamilton},
  {Hanley}, {Harch}, {Hart}, {Hersman}, {Hill}, {Hill}, {Hinson}, {Holdridge},
  {Horanyi}, {Howard}, {Howett}, {Jackman}, {Jacobson}, {Jennings}, {Kammer},
  {Kang}, {Kaufmann}, {Kollmann}, {Krimigis}, {Kusnierkiewicz}, {Lauer}, {Lee},
  {Lindstrom}, {Linscott}, {Lisse}, {Lunsford}, {Mallder}, {Martin}, {McComas},
  {McNutt}, {Mehoke}, {Mehoke}, {Melin}, {Mutchler}, {Nelson}, {Nimmo},
  {Nunez}, {Ocampo}, {Owen}, {Paetzold}, {Page}, {Parker}, {Parker},
  {Pelletier}, {Peterson}, {Pinkine}, {Piquette}, {Porter}, {Protopapa},
  {Redfern}, {Reitsema}, {Reuter}, {Roberts}, {Robbins}, {Rogers}, {Rose},
  {Runyon}, {Retherford}, {Ryschkewitsch}, {Schenk}, {Schindhelm}, {Sepan},
  {Showalter}, {Singer}, {Soluri}, {Stanbridge}, {Steffl}, {Strobel}, {Stryk},
  {Summers}, {Szalay}, {Tapley}, {Taylor}, {Taylor}, {Throop}, {Tsang},
  {Tyler}, {Umurhan}, {Verbiscer}, {Versteeg}, {Vincent}, {Webbert}, {Weidner},
  {Weigle}, {White}, {Whittenburg}, {Williams}, {Williams}, {Williams},
  {Woods}, {Zangari}, \& {Zirnstein}}]{Sternetal2015}
{Stern}, S.~A., {Bagenal}, F., {Ennico}, K., {et~al.} 2015,
  \href{http://dx.doi.org/10.1126/science.aad1815}{\JournalTitle{Science}, 350,
  aad1815}

\bibitem[{{Tegler} {et~al.}(2016){Tegler}, {Romanishin}, {Consolmagno}, \&
  {J.}}]{Tegleretal2016}
{Tegler}, S.~C., {Romanishin}, W., {Consolmagno}, G.~J., \& {J.}, S. 2016,
  \href{http://dx.doi.org/10.3847/0004-6256/152/6/210}{\JournalTitle{\aj}, 152,
  210}

\bibitem[{{Tsiganis} {et~al.}(2005){Tsiganis}, {Gomes}, {Morbidelli}, \&
  {Levison}}]{Tsiganisetal2005}
{Tsiganis}, K., {Gomes}, R., {Morbidelli}, A., \& {Levison}, H.~F. 2005,
  \href{http://dx.doi.org/10.1038/nature03539}{\JournalTitle{\nat}, 435, 459}

\bibitem[{{Volk} \& {Malhotra}(2008)}]{VolkMalhotra2008}
{Volk}, K., \& {Malhotra}, R. 2008,
  \href{http://dx.doi.org/10.1086/591839}{\JournalTitle{\apj}, 687, 714}

\bibitem[{{Volk} {et~al.}(2016){Volk}, {Murray-Clay}, {Gladman}, {Lawler},
  {Bannister}, {Kavelaars}, {Petit}, {Gwyn}, {Alexandersen}, {Chen}, {Lykawka},
  {Ip}, \& {Lin}}]{Volketal2016}
{Volk}, K., {Murray-Clay}, R., {Gladman}, B., {et~al.} 2016,
  \href{http://dx.doi.org/10.3847/0004-6256/152/1/23}{\JournalTitle{\aj}, 152,
  23}

\bibitem[{{Volk} {et~al.}(2017){Volk}, {Murray-Clay}, {Gladman}, {Chen}, {Lin},
  {Dawson}, {Lawler}, {Ip}, {Greenstreet}, {Lykawka}, {Alexandersen},
  {Bannister}, {Gwyn}, {Kavelaars}, \& {Petit}}]{VolkDPS2017}
{Volk}, K., {Murray-Clay}, R., {Gladman}, B., {et~al.} 2017, in AAS/Division
  for Planetary Sciences Meeting Abstracts, Vol.~49, AAS/Division for Planetary
  Sciences Meeting Abstracts, 216.12

\bibitem[{{Wong} \& {Brown}(2015)}]{WongBrown2015}
{Wong}, I., \& {Brown}, M.~E. 2015,
  \href{http://dx.doi.org/10.1088/0004-6256/150/6/174}{\JournalTitle{\aj}, 150,
  174}

\bibitem[{{Yoshida} \& {Terai}(2017)}]{YoshidaTerai2017}
{Yoshida}, F., \& {Terai}, T. 2017,
  \href{http://dx.doi.org/10.3847/1538-3881/aa7d03}{\JournalTitle{\aj}, 154,
  71}

\end{thebibliography}

 \newcommand{\noop}[1]{}

\end{document}